\documentclass[11pt,twoside]{article}

\usepackage{asp2004}
\usepackage{epsf}
\usepackage{psfig}
\usepackage{lscape}
\usepackage{natbib}
\usepackage{graphicx}

\begin{document}
\title{Pushing the limit on neutron star spin rates}   
\author{Duncan Galloway}
\affil{Centenary Fellow, School of Physics, University of Melbourne,
VIC 3010 Australia}

\begin{abstract} 
Millisecond X-ray pulsars consist of a rapidly-spinning neutron star
accreting from a low-mass stellar companion, and are the long-sought
evolutionary progenitors of millisecond radio pulsars,
as well as promising candidate sources for gravitational radiation.
The population
of these sources has grown significantly over the last three years, with
the discovery of six new examples
to bring the total sample to seven.
Three sources are ultracompact binaries with H-depleted donors and orbital
periods of $\approx40$~min, like the 185~Hz pulsar XTE~J0929$-$314.
Three more have orbital periods of 2~hr or longer, similar to
IGR~J00291+5934, first detected in outburst by 
{\it INTEGRAL} in December 2004.
The neutron star in this 2.46~hr binary has the most rapid
spin of the accreting pulsars at 599~Hz. 
The most recently-discovered pulsar, HETE~J1900.1$-$2455 (377~Hz), has an
intermediate orbital period of 83.3~min, and has been active 
for more than 1~yr, much longer than the typical transient outburst.
Pulsations were detected only in the first few months of the outburst;
this source has since resembled a faint, persistent non-pulsing low-mass
X-ray binary, typical of the broader low-mass X-ray binary population.
\end{abstract}

\section*{Introduction}

Neutron stars accreting from low-mass binary companions (low-mass X-ray
binaries, or LMXBs) have orbital periods of typically a few hours, and
characteristically exhibit
bursts from thermonuclear ignition of accreted material on the surface
\cite[e.g.][]{lew93}.
These objects are the evolutionary precursors to the ``recycled''
millisecond radio pulsars, which are thought to have been spun-up by an
extended period of accretion.  There are $>100$ LMXBs known
\cite[]{lmxb01}; most give no
observational indications of their (presumably) rapid spin,
although since 1996 some have been found to exhibit 
oscillations in the range 45--620~Hz only during thermonuclear bursts
\cite[e.g.][]{sb03}.
Since the discovery of the 401~Hz pulsar SAX~J1808.4$-$3658 in 1998
\cite[]{wij98b}, an additional subset of LMXBs
has emerged which consistently exhibit pulsations \cite[e.g.][]{wij03a}.
Two sources show pulsations and burst oscillations at the same frequency,
confirming that the latter phenomenon also traces the neutron-star spin
\cite[]{chak03a}.
Precisely what is different about these accretion-powered millisecond
pulsars (and the burst oscillation sources) that allows us to measure the
spin is not clear.

Here I describe the 
properties of three of the
seven accretion-powered millisecond pulsars, the first two 
representative of the two broad classes of binary which make up this
group.
Observations 
were made
with the Proportional Counter Array \cite[PCA;][]{xte96} aboard
the {\it Rossi X-ray Timing Explorer}\/ ({\it RXTE}).
The PCA consists of five co-aligned proportional counter units (PCUs), sensitive
to photons in the energy range 2--60~keV and with a total effective area
of $\approx6500\ {\rm cm^2}$. Arriving photons are time-tagged to
approximately $1\mu$s, and their energy is measured to a precision of
$<18$\% at 6~keV. The PCA is presently the only instrument with sufficient
sensitivity and temporal resolution to reliably detect the persistent
pulsations in the accretion-powered millisecond pulsars.

\section*{XTE \boldmath J0929$-$314}

This high--Galactic-latitude X-ray transient 
was first detected in outburst during April 2002 by the All-Sky Monitor
(ASM) aboard {\it RXTE}\/ \cite[]{gal02d}. The ASM consists of three
scanning shadow cameras sensitive to 1.5--12~keV photons which provides
90~s exposures of most of the sky every 96~min.
XTE~J0929$-$314 was identified using a ``deep sky map'' technique, in 
which maps of flux residuals are constructed by cross-correlating the
predicted coded mask pattern against the best-fit data residuals for
each $4'$ cell in the field of view.  
The resulting sensitivity for new sources is as low as $\simeq 15$~mCrab
away from the Galactic center, compared to the typical $\simeq 50$ mCrab
threshold for
individual camera snapshots. 

PCA observations revealed 185~Hz
(5.4~ms) pulsations, with a fractional rms amplitude of 3--7\%, modulated
by a 43.6~min binary orbit.  The Roche lobe in such a tiny binary cannot
contain a main-sequence companion, indicating that the mass donor is
highly evolved and H-poor \cite[]{nrj86}.
The system has one of the smallest measured mass functions ($2.7\times
10^{-7} M_\odot$) of any stellar binary.  The binary parameters imply a
$\simeq 0.01 M_\odot$ white dwarf donor with a moderately high
inclination. XTE~J0929$-$314 was the second ultracompact binary
millisecond pulsar discovered, after XTE~J1751$-$305
\cite[42~min;][]{markwardt02} and was later joined by a third,
XTE~J1807$-$294 \cite[40~min;][]{markwardt03b}. It is an open question why
the orbital periods of these sources should cluster so closely around
40~min.

XTE~J0929$-$314 remained X-ray active for just over two months, by which
time the {\it RXTE}\/ observations revealed spin-down at an average rate
of $\dot\nu=(-9.2\pm 0.4)\times 10^{-14}$ Hz~s$^{-1}$. The spin-down
torque may arise from magnetic coupling to the accretion disk, a
magnetohydrodynamic wind, or gravitational radiation from the rapidly
spinning pulsar. 

\section*{IGR J00291+5934}

This recurrent X-ray transient was initially identified by 
ESA's {\it International Gamma-Ray Astrophysics Laboratory}\/ 
{\it INTEGRAL}\/ in 2004 December, during a series of Galactic plane scans
conducted every 12~d \cite[]{shaw05}. The source was detected with the
IBIS/ISGRI coded-mask imager \cite[]{ibis03,isgri03}, which is sensitive
to photons in the range 15~keV to 1~MeV and with a
$29^\circ\times29^\circ$ field of view.

The {\it RXTE}/PCA observations which followed revealed 599~Hz (1.67 ms)
pulsations with a fractional rms amplitude of 9\% (3--13 keV).  The
neutron star is in a 2.46~hr orbit with a very low-mass donor, most likely
a brown dwarf heated by the X-ray emission originating from the neutron
star \cite[]{gal05a}.  The binary parameters of the system are similar to
those of the first known accreting millisecond pulsar, SAX J1808.4$-$3658.
While the companions of these systems are also very low-mass, like that of
XTE~J0929$-$314, the mass donors for the longer-orbital period systems are
likely still H-rich.

\begin{figure}
 \plottwo{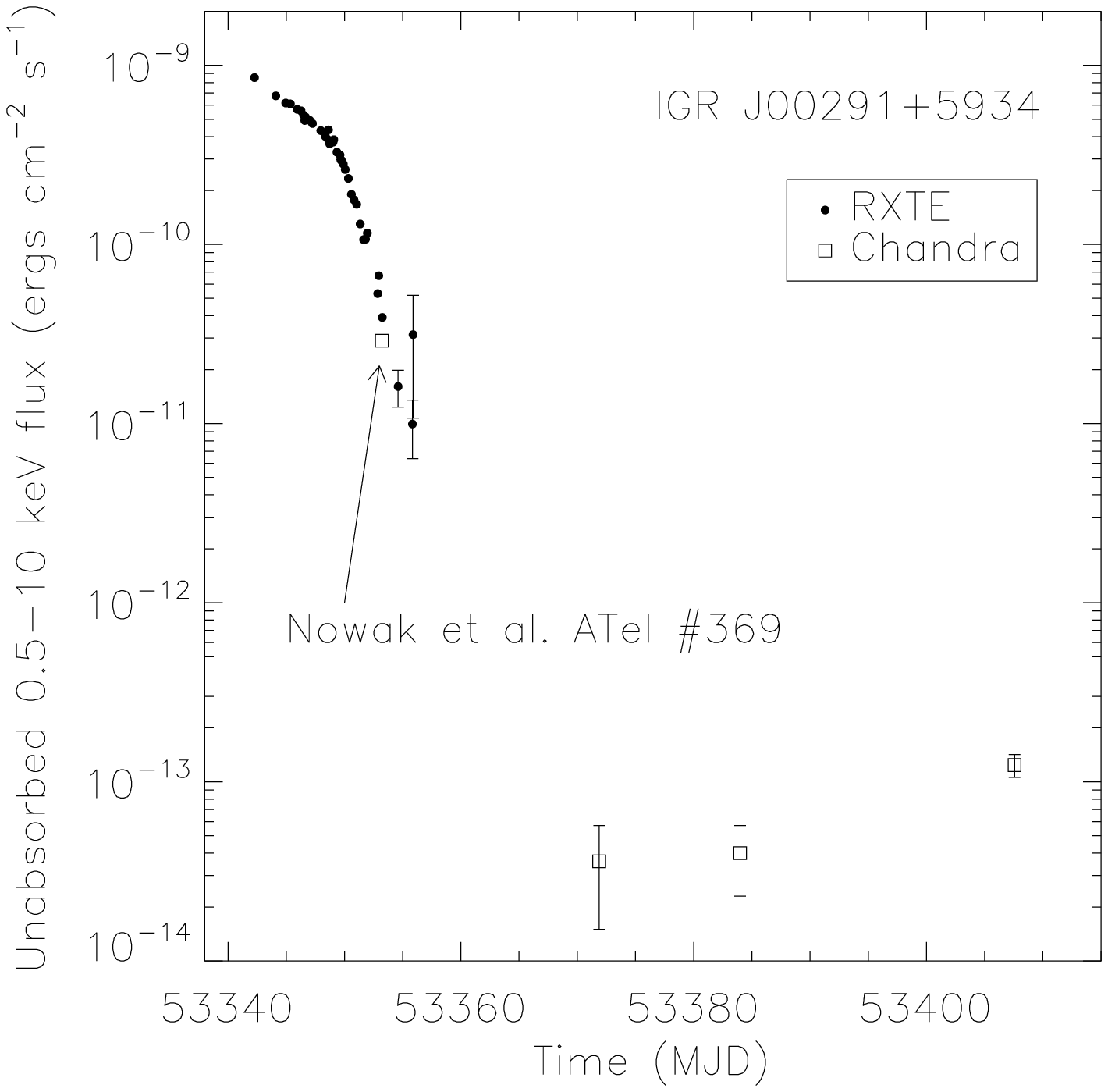}{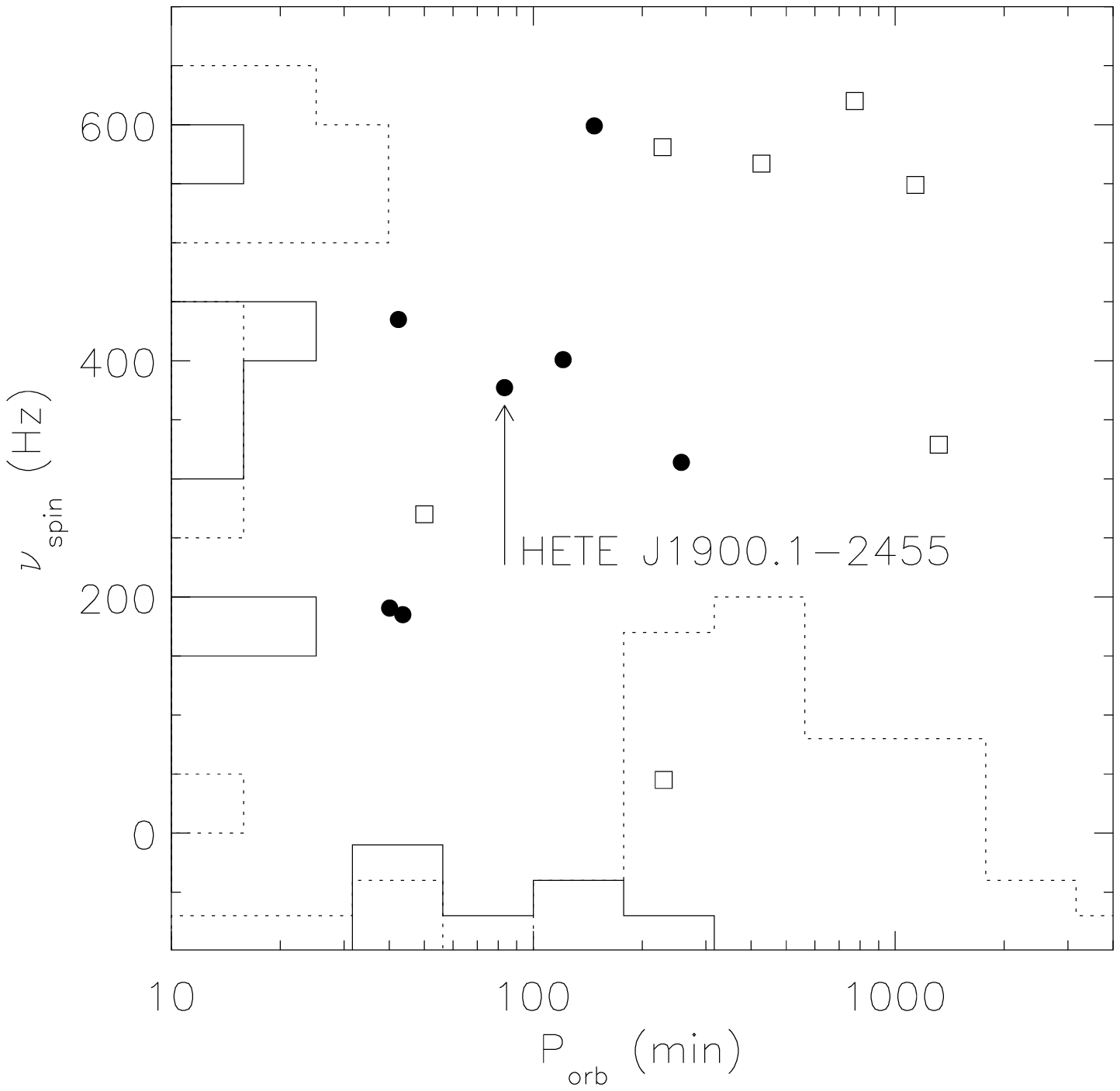} 
 \caption{{\it Left panel}\/ X-ray flux evolution of 
IGR~J00291+5934 during the 2004--5 outburst
and into quiescence. The inferred 0.5--10~keV flux from {\it RXTE}\/
observations is plotted as filled circles, while the flux from {\it
Chandra}\/ observations are plotted as open squares. After \cite{jonk05}.
{\it Right panel}\/
Spin frequency for rapidly-rotating accreting neutron stars as a
function of the binary orbital period. The orbital period is known for
7 of the 13 burst oscillation sources ({\it open squares}), but all 7 of
the known millisecond pulsars ({\it filled circles}). On each axis we also
plot the distribution of each parameter separately for the burst
oscillation sources ({\it dotted line}, including the sources without a
detected spin frequency) and pulsars ({\it solid line}).
  \label{fluxigr} }
\end{figure}

The source remained active for less than 20~d (Fig. \ref{fluxigr},
left).
{\it Chandra}\/ observations revealed significant variability
in quiescence \cite[]{jonk05}. Once the source position was known, a
search of earlier {\it RXTE}/ASM observations revealed two prior
outbursts, three and six years before. The outburst recurrence time is
similar to that of SAX~J1808.4$-$3658 \cite[see also][]{gal06b}.

\section*{HETE \boldmath J1900.1$-$2455}

The most recently-discovered accretion-powered
pulsar, HETE~J1900$-$2455 exhibits distinctly different behaviour in
several respects to the rest of the population.
The source was discovered in 2005 June when a strong thermonuclear
(type-I) burst was detected by {\it HETE-II}\/ \cite[]{vand05a}. {\it
HETE-II}\/ was designed to study gamma-ray bursts \cite[]{hete203}, but
thanks to sensitivity extending into the X-ray band can also detect
thermonuclear bursts from accreting neutron stars.

Subsequent {\it RXTE}\/ observations of the field 
revealed pulsations at 377.3~Hz
\cite[]{morgan05} as well as
Doppler shifts of the apparent pulsar frequency, originating from
orbital motion 
in the 83.25~min binary \cite[]{kaaret05b}. In this case the Roche lobe
can accommodate a brown dwarf with no need for extra heating.

The rms amplitude of the pulsations was unusually low compared to the
other pulsars, at best 2\%.  During the initial {\it RXTE}\/ observations,
the source underwent a short-lived flare after which the pulsations became
undetectable. Such behaviour has not been observed in the other six
pulsars, in which pulsations are consistently detected while the sources
are X-ray bright. However, the pulsar outburst has also lasted much longer
than in the other sources 
\cite[]{gal05d}, and in fact at the time of
writing (2006 June) the source is still active and being monitored with
regular {\it RXTE}\/ observations. Assuming that activity continues
at the present level, the time-averaged
accretion rate \cite[at the estimated distance of 5~kpc;][]{kawai05} is
the highest amongst all of the millisecond pulsars \cite[]{gal06b}.

\section*{Discussion}

Accretion-powered millisecond pulsars represent a rapidly-growing
subclass of LMXBs.
The present sample is sufficiently small that new examples can still
reveal vital clues as to the physics of these extreme objects;
consequently,
searches for, and observations of, new
examples are a high priority for {\it RXTE}\/ and other X-ray
missions.
Observations of HETE~J1900.1$-$2455 suggest, for the first time, a connection
between the transient pulsars and the non-pulsing (typically) persistent
LMXB population.

The combined spin distribution for rapidly-rotating neutron stars,
including both accretion- and rotation-powered, has a present maximum of
716~Hz \cite[]{hessels06}. This value is substantially below the expected
breakup frequency, and while selection effects may explain the
non-detection of faster radio pulsars, no such effects reduce the
likelihood of detecting faster pulsars in the X-ray band.  It has been
suggested that the torques which spin-up accreting neutron stars may be
balanced at high spin frequencies by gravitational radiation
\cite[e.g.][]{bil98c}.  A Bayesian analysis suggests that the observed
spin frequency distribution of accreting neutron stars is consistent with
a uniform distribution up to some maximum value between 700 and 800~Hz
\cite[]{chak03a,chak05a}. If this ``speed limit'' is indeed imposed by
gravitational wave emission, there is the exciting possibility of
detecting these sources with the next generation of gravitational wave
detectors, such as the Advanced Laser Interferometer Gravitational-Wave
Observatory \cite[LIGO; see e.g.][]{ligopulsar05}.
By virtue of its ``quasi-persistent'' behaviour, HETE~J1900.1$-$2455 is
presently the best accretion-powered millisecond pulsar candidate for the
detection of gravitational radiation.

While the observed distribution of spin frequencies does
not appear significantly different between the burst oscillation sources
and pulsars, the orbital periods are substantially shorter on average
(Fig. \ref{fluxigr}, right), particularly when compared to the entire LMXB
population. The K-S test statistic comparing the $P_{\rm orb}$
distribution of the millisecond pulsars and the known values for the
non-pulsing LMXBs \cite[]{lmxb01} is 0.751, corresponding to a 99.9\%
probability that the two populations are not drawn from the same
distribution.

The question of why certain sources exhibit persistent pulsations
and others do not, remains open. However, more detailed
studies of the behaviour of HETE~J1900.1$-$2455 may soon shed light on
this question.

\acknowledgements
This paper is based on collaborative work with Ed Morgan and Deepto
Chakrabarty at MIT; Tod Strohmayer and Craig Markwardt at NASA/GSFC; and
Phil Kaaret at U. Iowa.


\begin{thebibliography}{}
\expandafter\ifx\csname natexlab\endcsname\relax\def\natexlab#1{#1}\fi

\bibitem[{{Abbott}(2005)}]{ligopulsar05}
{Abbott}, B. et al. (LIGO Science Collaboration) 2005, \prl, 94, 181103

\bibitem[{{Bildsten}(1998)}]{bil98c}
{Bildsten}, L. 1998, \apjl, 501, L89

\bibitem[{{Chakrabarty}(2005)}]{chak05a}
{Chakrabarty}, D. 2005, in Binary Radio Pulsars, ed. F.~A. {Rasio} \& I.~H.
  {Stairs} (San Fransisco: ASP Conf. Ser. 328), 279 (astro--ph/0408004)

\bibitem[{{Chakrabarty} {et~al.}(2003){Chakrabarty}, {Morgan}, {Muno},
  {Galloway}, {Wijnands}, {van der Klis}, \& {Markwardt}}]{chak03a}
{Chakrabarty}, D., {Morgan}, E.~H., {Muno}, M.~P., {Galloway}, D.~K.,
  {Wijnands}, R., {van der Klis}, M., \& {Markwardt}, C.~B.  2003, \nat,
424, 42

\bibitem[{{Galloway}(2006)}]{gal06b}
{Galloway}, D.~K. 2006, in The Transient Milky Way: a perspective for MIRAX,
  ed. F.~{D'Amico}, J.~{Braga}, \& R.~{Rothschild} (Melville, NY: AIP;
  astro-ph/0604345)

\bibitem[{{Galloway} {et~al.}(2002){Galloway}, {Chakrabarty}, {Morgan}, \&
  {Remillard}}]{gal02d}
{Galloway}, D.~K., {Chakrabarty}, D., {Morgan}, E.~H., \& {Remillard}, R.~A.
  2002, \apjl, 576, L137

\bibitem[{{Galloway} {et~al.}(2005{\natexlab{a}}){Galloway}, {Markwardt},
  {Morgan}, {Chakrabarty}, \& {Strohmayer}}]{gal05a}
{Galloway}, D.~K., {Markwardt}, C.~B., {Morgan}, E.~H., {Chakrabarty}, D.,
\& {Strohmayer}, T.~E.  2005{\natexlab{a}}, \apjl, 622, L45

\bibitem[{{Galloway} {et~al.}(2005{\natexlab{b}}){Galloway}, {Morgan},
  {Kaaret}, {Chakrabarty}, \& {Suzuki}}]{gal05d}
{Galloway}, D.~K., 
 {Morgan}, E.~H., {Kaaret}, P., {Chakrabarty}, D., \& {Suzuki}, M.
  2005{\natexlab{b}}, The Astronomer's Telegram, 657

\bibitem[{{Hessels} {et~al.}(2006){Hessels}, {Ransom}, {Stairs}, {Freire},
  {Kaspi}, \& {Camilo}}]{hessels06}
{Hessels}, J. W.~T., {Ransom}, S.~M., {Stairs}, I.~H., {Freire}, P. C.~C., {Kaspi}, V.~M., \& {Camilo}, F.
  2006, Science, in press (astro-ph/0601337)

\bibitem[{{Jahoda} {et~al.}(1996){Jahoda}, {Swank}, {Giles}, {Stark},
  {Strohmayer}, {Zhang}, \& {Morgan}}]{xte96}
{Jahoda}, K.,  {Swank}, J.~H., {Giles}, A.~B., {Stark}, M.~J., {Strohmayer}, T.,
  {Zhang}, W., \& {Morgan}, E.~H.
  1996, Proc. SPIE, 2808, 59

\bibitem[{{Jonker} {et~al.}(2005){Jonker}, {Campana}, {Steeghs}, {Torres},
  {Galloway}, {Markwardt}, {Chakrabarty}, \& {Swank}}]{jonk05}
{Jonker}, P.~G., 
 {Campana}, S., {Steeghs}, D., {Torres}, M.~A.~P., {Galloway},
  D.~K., {Markwardt}, C.~B., {Chakrabarty}, D., \& {Swank}, J.
  2005, \mnras, 361, 511

\bibitem[{{Kaaret} {et~al.}(2006){Kaaret}, {Morgan}, {Vanderspek}, \&
  {Tomsick}}]{kaaret05b}
{Kaaret}, P., {Morgan}, E.~H., {Vanderspek}, R., \& {Tomsick}, J.~A. 2006,
  \apj, 638, 963

\bibitem[{{Kawai} {et~al.}(2005){Kawai}, {Suzuki}, \& {for the HETE
  Team}}]{kawai05}
{Kawai}, N., {Suzuki}, M., \& {for the HETE Team}. 2005, The Astronomer's
  Telegram, 534

\bibitem[{{Lebrun} {et~al.}(2003){Lebrun}, {Leray}, {Lavocat}, {Cr{\'e}tolle},
  {Arqu{\`e}s}, {Blondel}, {Bonnin}, {Bou{\`e}re}, {Cara}, {Chaleil}, {Daly},
  {Desages}, {Dzitko}, {Horeau}, {Laurent}, {Limousin}, {Mathy}, {Mauguen},
  {Meignier}, {Molini{\'e}}, {Poindron}, {Rouger}, {Sauvageon}, \&
  {Tourrette}}]{isgri03}
{Lebrun}, F. et al.
  2003, \aap, 411, L141

\bibitem[{{Lewin} {et~al.}(1993){Lewin}, {van Paradijs}, \& {Taam}}]{lew93}
{Lewin}, W. H.~G., {van Paradijs}, J., \& {Taam}, R.~E. 1993, \ssr, 62, 223

\bibitem[{{Liu} {et~al.}(2001){Liu}, {van Paradijs}, \& {van den
  Heuvel}}]{lmxb01}
{Liu}, Q.~Z., {van Paradijs}, J., \& {van den Heuvel}, E.~P.~J. 2001, \aap,
  368, 1021

\bibitem[{{Markwardt} {et~al.}(2003){Markwardt}, {Juda}, \&
  {Swank}}]{markwardt03b}
{Markwardt}, C.~B., {Juda}, M., \& {Swank}, J.~H. 2003, IAU Circ.

\bibitem[{{Markwardt} {et~al.}(2002){Markwardt}, {Swank}, {Strohmayer}, {in 't
  Zand}, \& {Marshall}}]{markwardt02}
{Markwardt}, C.~B., {Swank}, J.~H., {Strohmayer}, T.~E., {in 't Zand}, J.
  J.~M., \& {Marshall}, F.~E. 2002, \apjl, 575, L21

\bibitem[{{Morgan} {et~al.}(2005){Morgan}, {Kaaret}, \&
  {Vanderspek}}]{morgan05}
{Morgan}, E., {Kaaret}, P., \& {Vanderspek}, R. 2005, The Astronomer's
  Telegram, 523

\bibitem[{{Nelson} {et~al.}(1986){Nelson}, {Rappaport}, \& {Joss}}]{nrj86}
{Nelson}, L.~A., {Rappaport}, S.~A., \& {Joss}, P.~C. 1986, \apj, 304, 231

\bibitem[{{Ricker} {et~al.}(2003){Ricker}, {Atteia}, {Crew}, {Doty},
  {Fenimore}, {Galassi}, {Graziani}, {Hurley}, {Jernigan}, {Kawai}, {Lamb},
  {Matsuoka}, {Pizzichini}, {Shirasaki}, {Tamagawa}, {Vanderspek}, {Vedrenne},
  {Villasenor}, {Woosley}, \& {Yoshida}}]{hete203}
{Ricker}, G.~R. et al.
  2003, in AIP Conf. Proc. 662: Gamma-Ray
  Burst and Afterglow Astronomy 2001: A Workshop Celebrating the First Year of
  the HETE Mission, ed. G.~R. {Ricker} \& R.~K. {Vanderspek}, 3--16

\bibitem[{{Shaw} {et~al.}(2005){Shaw}, {Mowlavi}, {Rodriguez}, {Ubertini},
  {Capitanio}, {Ebisawa}, {Eckert}, {Courvoisier}, {Produit}, {Walter}, \&
  {Falanga}}]{shaw05}
{Shaw}, S.~E. et al.
  2005, \aap, 432, L13

\bibitem[{{Strohmayer} \& {Bildsten}(2006)}]{sb03}
{Strohmayer}, T. \& {Bildsten}, L. 2006, in Compact Stellar X-Ray Sources, ed.
  W.~H.~G. {Lewin} \& M.~{van der Klis} (Cambridge University Press),
  (astro--ph/0301544)

\bibitem[{{Ubertini} {et~al.}(2003){Ubertini}, {Lebrun}, {Di Cocco}, {Bazzano},
  {Bird}, {Broenstad}, {Goldwurm}, {La Rosa}, {Labanti}, {Laurent}, {Mirabel},
  {Quadrini}, {Ramsey}, {Reglero}, {Sabau}, {Sacco}, {Staubert}, {Vigroux},
  {Weisskopf}, \& {Zdziarski}}]{ibis03}
{Ubertini}, P. et al.
  2003, \aap, 411, L131

\bibitem[{{Vanderspek} {et~al.}(2005){Vanderspek}, {Morgan}, {Crew},
  {Graziani}, \& {Suzuki}}]{vand05a}
{Vanderspek}, R., {Morgan}, E., {Crew}, G., {Graziani}, C., \& {Suzuki}, M.
  2005, The Astronomer's Telegram, 516

\bibitem[{{Wijnands}(2004)}]{wij03a}
{Wijnands}, R. 2004, in Proceedings of the 2nd BeppoSAX Conference: "The
  Restless High-Energy Universe", Amsterdam, 5--9 May 2003, ed. E.~P.~J. {van
  den Heuvel}, R.~A.~M.~J. {Wijers}, \& J.~J.~M. {in 't Zand}, Vol. 132,
  496--505

\bibitem[{{Wijnands} \& {van der Klis}(1998)}]{wij98b}
{Wijnands}, R. \& {van der Klis}, M. 1998, \nat, 394, 344

\end{thebibliography}
\end{document}